\documentclass[twocolumn,showpacs,pre,floatfix]{revtex4}
\usepackage{amsmath,amsfonts,amssymb}
\usepackage{graphicx} 

\begin{document}
\title{Granular discorectangle in a thermalized bath of hard disks}
\author{H. Gomart$^1$}
\author{J. Talbot$^2$} 
\author{P. Viot$^1$}

\affiliation{$^1$Laboratoire de Physique
Th\'eorique des Liquides, Universit\'e   Pierre et Marie Curie,  4,  place
Jussieu, 75252 Paris Cedex, 05 France}
 \affiliation{$^2$Department   of  Chemistry  and  Biochemistry,
 Duquesne University, Pittsburgh, PA 15282-1530}

\begin{abstract}
By  using  the  Enskog-Boltzmann approach,  we  study the steady-state
dynamics of a granular discorectangle placed in a two-dimensional bath
of  thermalized hard disks.  Hard  core collisions are assumed elastic
between disks and inelastic  between the discorectangle and the disks,
with  a normal  restitution  coefficient $\alpha<1$.   Assuming a  Gaussian
ansatz   for  the probability     distribution  functions,  we  obtain
analytical  expressions for  the  granular temperatures.  We show  the
absence   of equipartition  and   investigate  both  the  role of  the
anisotropy of the discorectangle and of the relative ratio of the bath
particles  to the linear sizes of  the discorectangle. In addition, we
investigate a  model of a discorectangle   with two normal restitution
coefficients for collisions along the  straight and curved surfaces of
the discorectangle. In this case  one observes equipartition for a non
trivial ratio of normal restitution coefficients.
\end{abstract}
\maketitle
\section{Introduction}
Granular    matter is characterized  by   the existence of dissipative
forces between particles. In order to  sustain a collective motion, it
is  necessary to provide   energy continuously.  When power supply  is
sufficiently copious, the assembly of granular particles attains a non
equilibrium steady state  (NESS)\cite{NL98,MP99}.  It  is customary to
define the granular  temperature as the second  moment of the velocity
distribution.  It is a   source of both fascination and  inconvenience
that  the well-know properties  of temperatures characterizing thermal
systems are not necessarily transferable  to granular temperatures. In
particular,                 recent        work,                   both
theoretical\cite{GD99,MP99,BT02,DHGD02,CH02,MG03,AL03}             and
experimental\cite{FM02,WP02},  has shown  that  in  a binary  granular
system the two species  have different granular temperatures  that are
non-trivial functions   of the  microscopic  parameters  (mass,  size,
restitution coefficient,\ldots).  Although  the absence of equipartition is
not surprising  for a dissipative system sustained  in  a NESS, a more
complete  investigation is  necessary  since the granular temperatures
play an important  role in hydrodynamic  descriptions of these systems
(in particular the absence  of equipartition in binary mixtures yields
granular  temperature gradients which enhance segregation\cite{DH04}).
In addition, the extension of  the fluctutation-dissipation theorem is
an important issue in the context of granular gases\cite{BLP04}. Other
consequences of the absence of equipartition  include the ability of a
binary  system  to  exhibit a segregation   phenomena   in a ``Maxwell
demon'' experiment\cite{BT03}.  See also the homogeneous cooling state
of a granular mixture\cite{SD01} and the impurity problem\cite{SD01b}.

Most of the above-referenced  studies examined assemblies of spherical
particles. Yet, in  reality, the particles  composing granular systems
are to some  degree anisotropic and, in many  cases, strongly so. Even
if the  particles are smooth,  each collision results in some exchange
and, possibly loss, of rotational kinetic energy. There are relatively
few   studies    of   these     systems   (but    see  for    example,
Refs.\cite{P95,AT03,M04})  and fewer still  that focus specifically on
equipartition.   Huthmann et al.  \cite{HAZ99}  used kinetic theory to
examine  the free  cooling of a  system of  granular needles  in three
dimensions and, more  recently, two of  the present authors studied  a
two-dimensional  system composed of   a single  granular needle  in  a
thermalized  bath of   point  particles \cite{VT04}  in  a  NESS.  For
inelastic needle-point collisions, the rotational granular temperature
is smaller  than the  translational one while  both are  less than the
bath  temperature.  The validity  of the  theoretical predictions were
confirmed by comparison with numerical simulations of the model. While
this   study provided  useful insights,    infinitesimal width of  the
particle is obviously an idealization.

The  objective of  present article   is to consider  a more  realistic
system where both the tracer particle and bath particles are of finite
extent.  Specifically, we  consider a  discorectangle   in a  bath  of
thermalized hard disks. Fortunately, despite the increased complexity,
it is  still possible  to obtain  an analytic  solution of the  steady
state kinetic   equations.  The  principal   difference between    the
discorectangle-disk and needle-point   systems  is that two   kinds of
collision  are  possible  in  the  former   compared   to one  in  the
latter. Specifically, a disk can collide  with either the sides or the
caps of the discorectangle. If each type of collision is characterized
by     different  normal  restitution    coefficients   we show   that
equipartition   between the  translational   and rotational degrees of
freedom    can    be  obtained   for      specific  values  of   these
parameters. Consequently,  for appropriate ranges  of the  restitution
coefficients the translational  granular temperature may  be less than
or greater than the rotational one.

\section{Model and collision rules}
\begin{figure}
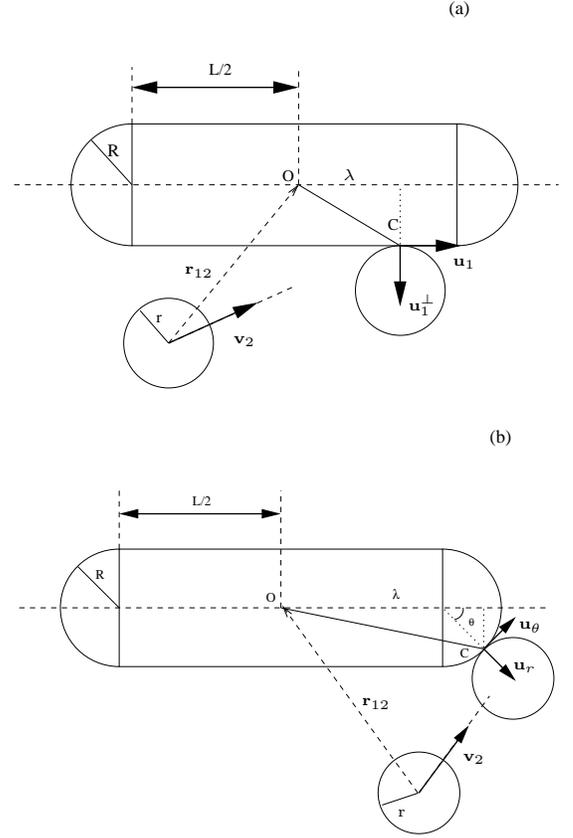

\centering
\begin{picture}(0,0)%
\includegraphics{discorect.pstex}%
\end{picture}%
\setlength{\unitlength}{2238sp}%
\begingroup\makeatletter\ifx\SetFigFont\undefined%
\gdef\SetFigFont#1#2#3#4#5{%
  \reset@font\fontsize{#1}{#2pt}%
  \fontfamily{#3}\fontseries{#4}\fontshape{#5}%
  \selectfont}%
\fi\endgroup%
\begin{picture}(6054,4306)(979,-3671)
\put(4636,-1366){\makebox(0,0)[lb]{\smash{{\SetFigFont{6}{7.2}{\rmdefault}{\mddefault}{\updefault}$\lambda$}}}}
\put(3421,-3166){\makebox(0,0)[lb]{\smash{{\SetFigFont{6}{7.2}{\rmdefault}{\mddefault}{\updefault}${\bf v}_2$}}}}
\put(5851,-2311){\makebox(0,0)[lb]{\smash{{\SetFigFont{6}{7.2}{\rmdefault}{\mddefault}{\updefault}${\bf u}_1$}}}}
\put(5356,-2806){\makebox(0,0)[lb]{\smash{{\SetFigFont{6}{7.2}{\rmdefault}{\mddefault}{\updefault}${\bf u}_1^{\perp}$}}}}
\put(2881,-2401){\makebox(0,0)[lb]{\smash{{\SetFigFont{6}{7.2}{\rmdefault}{\mddefault}{\updefault}${\bf r}_{12}$}}}}
\end{picture}%
\\[0.5cm]
\begin{picture}(0,0)%
\includegraphics{discorect2.pstex}%
\end{picture}%
\setlength{\unitlength}{2196sp}%
\begingroup\makeatletter\ifx\SetFigFont\undefined%
\gdef\SetFigFont#1#2#3#4#5{%
  \reset@font\fontsize{#1}{#2pt}%
  \fontfamily{#3}\fontseries{#4}\fontshape{#5}%
  \selectfont}%
\fi\endgroup%
\begin{picture}(6054,4582)(1159,-3947)
\put(6744,-2085){\makebox(0,0)[lb]{\smash{{\SetFigFont{6}{7.2}{\rmdefault}{\mddefault}{\updefault}${\bf u}_r$}}}}
\put(6811,-1621){\makebox(0,0)[lb]{\smash{{\SetFigFont{6}{7.2}{\rmdefault}{\mddefault}{\updefault}${\bf u}_{\theta}$}}}}
\put(6180,-3113){\makebox(0,0)[lb]{\smash{{\SetFigFont{6}{7.2}{\rmdefault}{\mddefault}{\updefault}${\bf v}_2$}}}}
\put(5446,-3751){\makebox(0,0)[lb]{\smash{{\SetFigFont{6}{7.2}{\rmdefault}{\mddefault}{\updefault}r}}}}
\put(5041,-2491){\makebox(0,0)[lb]{\smash{{\SetFigFont{6}{7.2}{\rmdefault}{\mddefault}{\updefault}${\bf r}_{12}$}}}}
\end{picture}%

\caption{Geometry of the discorectangle  and a disk in the plane: (a)
Collision   between the rectilinear part of   the discorectangle and a
disk, ${\bf r}_{12}$  denotes a vector  joining the  point labeled $2$
and the center  of the discorectangle;  ${\bf u}_1$  is a unit  vector
along the  axis of the discorectangle, $\lambda$  is  the algebraic distance
between the center  of the discorectangle and the  point of impact and
${\bf   u}_1^\perp$ is a  unit  vector  perpendicular to  the  axis of the
discorectangle.  (b) Collision   between   the circular part  of   the
discorectangle and a disk. $\theta$ denotes the angle between the direction
of the long  axis of the  discorectangle and the  axis  defined by the
contact point and  the center of  the disk.${\bf u}_{\bf r}$ and ${\bf
u}_{\theta }$ are unit   vectors normal and tangential to   the  surface at the  point  of
contact, respectively. }
\label{fig:1} 
\end{figure}

We investigate a two-dimensional system consisting of a discorectangle
of total  length  $L+2R$, radius  $R$  and mass $M$   with a moment of
inertia $I$ (The  value  of which is  given  in appendix A).  The bath
consists of disks of mass $m$ and of radius $r$.  The vector positions
of the center  of mass of the  discorectangle and a  disk particle are
denoted by ${\bf r}_1$ and ${\bf  r}_2$, respectively. The orientation
of the discorectangle  is specified by a  unit vector ${\bf u}_1$ that
points  along the long  axis.  Let  ${\bf r}_{12}={\bf r}_1-{\bf r}_2$
and  ${\bf u}_{1}^\perp$ denote a vector  perpendicular to ${\bf u}_1$.  A
collision between a discorectangle and a disk can take place either on
the linear part or on the circular parts of the former,
\begin{equation}\label{eq:1}
{\bf r}_{12}.{\bf u}_{1}^\perp=-(R+r),
\end{equation}
if  $|\lambda|<L/2$, and
\begin{equation}
{\bf r}_{12}.{\bf u}_r=-(R+r),
\end{equation}
if $R>|\lambda|-L/2>0$, where ${\bf u}_r$ denotes  the unit  vector of the
collision axis.  (see  Fig.~\ref{fig:1}). The relative velocity of the
point of contact ${\bf V}$ is given by
\begin{equation}\label{eq:2}
{\bf V}={\bf v}_{12}+{{\boldsymbol \omega}_1}\times {\bf OC},
\end{equation}
where ${\boldsymbol\omega_1}$  denotes the angular time derivative and
${\bf OC}$ the vector from the center of the  discorectangle
to the  point of impact.

The pre- and post-collisional quantities (the latter are labeled with
a prime) satisfy:
\begin{itemize}			
\item Total momentum conservation
\begin{equation}\label{eq:3}
M {\bf v'}_{1}+m{\bf v'}_2=M {\bf v}_{1}+m{\bf v}_2.
\end{equation}
\item Angular momentum conservation with respect to the point of
contact
\begin{equation}\label{eq:4}
I\omega'_1{\bf k}=I\omega_1{\bf k}+m{\bf OC}\times \left({\bf v}_{2}-{\bf v'}_{2}\right),
\end{equation}
where ${\bf k}$ is a unit vector perpendicular to the plane.
\end{itemize}
As a result of the collision, the  relative velocity of the contacting
points changes instantaneously according to the following relations:
\begin{align}\label{eq:5}
{\bf V'}.{\bf u}_1^\perp &=-\alpha{\bf V}.{\bf u}_1^\perp, \\
{\bf V'}.{\bf u}_1 &={\bf V}_.{\bf u}_1, \label{eq:6}
\end{align}
where $\alpha$   is   the normal restitution   coefficient and
 ${\bf u}_1^\perp$ and ${\bf u}_1$  denote the unit vectors of the
collision along the rectilinear part of the discorectangle.
When the collision occurs on the circular parts of the discorectangle,
the collision rules are given by
\begin{align}\label{eq:7}
{\bf V'}.{\bf u}_r &=-\alpha{\bf V}.{\bf u}_r, \\
{\bf V'}.{\bf u}_\theta &={\bf V}_.{\bf u}_\theta, \label{eq:8}
\end{align}
where ${\bf u}_r$ and ${\bf u}_\theta$ denote the unit vectors associated with
the circular part of the discorectangle (See Fig~\ref{fig:1}b).

The tangential restitution coefficient  is set to  one for the sake of
simplicity.    This  choice   is    reflected    in    the   form   of
Eqs.~(\ref{eq:6}-\ref{eq:8}).

By  combining Eqs.~(\ref{eq:2})-~(\ref{eq:5})  one obtains, after some
algebra, the change of the  discorectangle momentum $\Delta {\bf p}=M ({\bf
v}'_1- {\bf v}_1)$ for a collision along the linear part
\begin{equation}\label{eq:9}
\Delta{\bf          p}.{\bf        u}_{1}^\perp=-\frac{(1+\alpha){\bf        V}.{\bf
u}_{1}^\perp}{\frac{1}{m}+\frac{1}{M}+\frac{\lambda^2}{I}}.
\end{equation}
for $|\lambda |\leq L/2$
 and at the two ends of the discorectangle
\begin{equation}
\Delta{\bf          p}.{\bf        u}_r=-\frac{(1+\alpha){\bf        V}.{\bf
u}_r}{\frac{1}{m}+\frac{1}{M}+\frac{L^2\sin^2\theta }{4I}}.
\end{equation}
for  $R>|\lambda|-L/2>0$ with $\cos(\theta)=\frac{\lambda-L/2}{R}$.

\section{Boltzmann equation}
Since  we  are  interested  in  the homogeneous state,   the
distribution  function $f({\bf v}_1,\omega_1)$ of  the  discorectangle obeys
\begin{equation}\label{eq:10}
\frac{\partial f( {\bf v}_1,\omega_1)}{\partial t}=N \int\frac{d\theta_1}{2\pi}\int d{\bf v_2}\int d{\bf r}_2
\overline{ T}_{12}f( {\bf v}_1,\omega_1,{\bf v}_2)
\end{equation} 
where $N$ is the total number of disks, $f( {\bf v}_1,\omega_1,{\bf v}_2)$ is
the distribution function of the discorectangle and a disk, and $\overline{
T}_{12}$ is the collision operator between a discorectangle and a disk.

Defining the granular temperatures as quadratic average of the appropriate
velocity distribution, one has $T_T=M/2\langle  {\bf v}^2\rangle $ and
$T_R=I\langle \omega^2 \rangle$ for the translational and rotational granular
temperatures,  respectively (the angular brackets denote the average). By
taking the second moment with respect of the velocity and of the
angular velocity of Eq.(\ref{eq:10}), one obtains
\begin{align}\label{eq:11}
\frac{2\partial T_T}{M\partial t} &=\int d{\bf v}_1d{ \omega}_1\partial_t({\bf
v}_1^2f({\bf v}_1,\omega_1))\nonumber\\
=N&\int d{\bf v}_1\int d{ \omega}_1 \int\frac{d\theta_1}{2\pi}\int d{\bf v}_2\int d{\bf r}_2\overline{T}_{12}f({\bf v}_1,\omega_1,{\bf v}_2){\bf v}_1^2,
\end{align}
\begin{align}\label{eq:12}
\frac{\partial T_R}{I\partial t} &=\int d{\bf v}_1d{
\omega}_1\partial_t({ \omega}_1^2f({\bf v}_1,\omega_1))\nonumber\\
 =N&\int d{\bf v}_1\int d{ \omega}_1\int \frac{d\theta_1}{2\pi}\int d{\bf v}_2\int d{\bf r}_2\overline{T}_{12}f({\bf v}_1,\omega_1,{\bf v}_2){ \omega}_1^2.
\end{align}
In the stationary state the
time derivatives of the left-hand side of these two equations are
equal to zero.

To build the collision operator between the discorectangle and a disk,
$\overline{ T}_{12}$,  one   must  include the change   in  quantities
(i.e. velocity and angular momentum) produced during the infinitesimal
time interval of the collision.  This  operator is different from zero
only if the  two particles are in   contact and if  the particles were
approaching just  before   the collision\cite{AHZ00}. For   a collision
between a disk and  the  rectilinear part  of the discorectangle,  the
explicit form of the operator given by
\begin{align}\label{eq:13}
\overline{ T}_{12}&\propto \Theta(L/2-|\lambda|)\delta(|{\bf r}_{12}.{\bf u}_{1}^\perp|-r-R) \nonumber\\
&\times \left|\frac{d|{\bf r}_{12}.{\bf u}_{1}^\perp|}{dt}\right|
\Theta \left(-\left|\frac{d|{\bf r}_{12}.{\bf u}_{1}^\perp|}{dt}\right|\right)(b_{12}-1),
\end{align}
where $b_{12}$ is an operator  that changes post-collisional quantities
to pre-collisional quantities and $\Theta(x)$ is the Heaviside function,
and for a collision at two ends of the discorectangle
\begin{align}
\overline{ T}_{12}&\propto \Theta(R+L/2-|\lambda|)\Theta(|\lambda|-L/2)  \delta(|{\bf r}_{12}.{\bf u}_r|-(r+R))\nonumber\\
&\times \left|\frac{d|{\bf r}_{12}.{\bf u}_r|}{dt}\right|
\Theta \left(-\left|\frac{d|{\bf r}_{12}.{\bf u}_r|}{dt}\right|\right)(b_{12}-1),
\end{align}
where $b_{12}$ is an operator that converts pre-collisional 
to post-collisional quantities.

The others terms of the collision operator correspond to the necessary
conditions      of   contact    $ \Theta   (L/2-|\lambda|)\delta(|{\bf    r}_{12}.{\bf
u}_{1}^\perp|-(r+R))$,    and  approach        $\Theta\left(-\left|\frac{d|{\bf
r}_{12}.{\bf u}_{1}^\perp|}{dt}\right|\right)$ in the  first case and  $ \Theta(R+L/2-|\lambda|)\Theta(|\lambda|-L/2)\delta(|{\bf    r}_{12}.{\bf  u}_r|-(r+R))$,   and    approach
$\Theta\left(-\left|\frac{d|{\bf r}_{12}.{\bf u}_r|}{dt}\right|\right)$  in
the second case.

By taking  the  second moments  of  the  distribution function of  the
discorectangle   and  after  substitution    of     the  collision    operator
(Eq.(\ref{eq:13})),  one   obtains explicitly  for   the translational
kinetic energy

\begin{align}\label{eq:14}
&\int...\int  d{\bf r}_2 d{\bf v}_1 d{\bf v}_2 d\omega_1 d\theta_1 \left[ \Theta(L/2-|\lambda|)\delta(|{\bf r}_{12}.{\bf u}_{1}^\perp|-(r+R))\right.\nonumber\\
&\times \left|\frac{d|{\bf r}_{12}.{\bf u}_{1}^\perp|}{dt}\right|
\Theta\left(-\left|\frac{d|{\bf r}_{12}.{\bf
u}_{1}^\perp|}{dt}\right|\right)
f( {\bf v}_1, \omega_1)\Phi({\bf v}_2)  \Delta E^T_1 \nonumber\\&+  \Theta(R+L/2-|\lambda|)\Theta(|\lambda|-L/2)\delta(|{\bf r}_{12}.{\bf u}_r|-(r+R))\nonumber\\
&\times \left.\left|\frac{d|{\bf r}_{12}.{\bf u}_r|}{dt}\right|
\Theta\left(-\left|\frac{d|{\bf r}_{12}.{\bf
u}_r|}{dt}\right|\right)
f( {\bf v}_1,{\bf \omega}_1)\Phi({\bf v}_2)  \Delta E^T_1\right]
=0.
\end{align}

A similar equation can be written for rotational kinetic energy. Since
the impulse of the collision depends on the location of the impact, it
is easy  to  show that the  solution  is  not a   Maxwell distribution
function, a property already observed  in  the model  of a needle  and
points\cite{VT04}.   However, since the  deviations from the Maxwell distribution  are
small, we use it as a trial function. That is we assume that
\begin{equation}\label{eq:15}
f({\bf v}_1, { \omega}_1)\propto \exp\left(-\frac{M{\bf v_1}^2}{2\gamma_TT}-\frac{I{ \omega}_1^2}{2\gamma_RT}\right),
\end{equation}
where  $\gamma_T$  and   $\gamma_R$ are  the  ratios of   the  translational and
rotational discorectangle temperatures to the bath temperature, respectively.
In summary, in order to obtain the granular temperatures of the
discorectangle, it is necessary to: (i) Calculate the
change of translational and rotational energy occurring during a
collision  (ii) Perform an
average over all degrees of freedom of the collision integral.
\section{Calculation and Results}
\subsection{Energy changes during a collision}

When a   disk   collides with the    discorectangle, the   change   of  the
translational kinetic energy of the latter is given by
\begin{align}\label{eq:16}
 \Delta E_{1}^T&=\frac{M}{2}\left(({\bf v}'_{1})^2-({\bf v}_{1})^2\right)\nonumber\\
&=\Delta {\bf p}.{\bf v}_{1}+\frac{1}{M}\frac{\Delta {\bf p}^2}{2}\nonumber\\
&=-(1+\alpha)\frac{{\bf V}.{\bf u}_{1}^\perp{\bf v}_1.{\bf u}_{1}^\perp}
{\frac{1}{m}+\frac{1}{M}+\frac{\lambda^2}{I}}+\frac{1}{2M}
\frac{(1+\alpha)^2({\bf V}.{\bf u}_{1}^\perp)^2}
{\left(\frac{1}{m}+\frac{1}{M}+\frac{\lambda^2}{I}\right)^2},
\end{align}
for $\vert \lambda \vert < \frac L2$,  and 
\begin{align}\label{eq:17}
\Delta E_1^T &=-\frac{(1+\alpha){\bf V}.{\bf u}_r{\bf v}_1.{\bf
u}_r}{\frac1m+\frac1M+\frac{L^2\sin^2\theta}{4I}}
\nonumber\\
&+\frac{1}{2M}\frac{(1+\alpha)^2({\bf                               V}.{\bf
u}_r)^2}{(\frac1m+\frac1M+\frac{L^2\sin^2\theta}{4I})^2}.
\end{align}
for $R>\vert \lambda \vert - \frac L2>0$. 

The collision results in  a change of rotational energy, for $|\lambda|<\frac L2$,
\begin{align}
\Delta E_1^R&=\frac{I}{2}\left((\omega'_1)^2-(\omega_1)^2\right)\nonumber\\
&=-\frac{\lambda(1+\alpha)}{2}\frac{{\bf V}.{\bf u}_{1}^\perp(\omega_1'+\omega_1)}
{\frac{1}{m}+\frac{1}{M}+\frac{\lambda^2}{I}}\nonumber\\
&=-\lambda(1+\alpha)\frac{{\bf V}.{\bf u}_{1}^\perp\omega_1}
{\frac{1}{m}+\frac{1}{M}+\frac{\lambda^2}{I}}+
\frac{\lambda^2(1+\alpha)^2({\bf V}.{\bf u}_{1}^\perp)^2}
{2I\left(\frac{1}{m}+\frac{1}{M}+\frac{\lambda^2}{I}\right)^2}.
\end{align}
and for  $R>|\lambda|-L/2>0$,
\begin{align}
\Delta E_1^R&=-(1+\alpha)\frac{L\sin \theta{\bf V}.{\bf u}_r\omega_1}
{2\left(\frac{1}{m}+\frac{1}{M}+\frac{L^2\sin^2\theta}{4I}\right)}\nonumber\\&+
\frac{(1+\alpha)^2L^2\sin^2 \theta({\bf V}.{\bf u}_r)^2}
{8I\left(\frac{1}{m}+\frac{1}{M}+\frac{L^2\sin^2\theta}{4I}\right)^2}.
\end{align} 

\subsection{Expressions of the granular temperatures}
\begin{figure}[ht]
\resizebox{8.0cm}{!}{\includegraphics{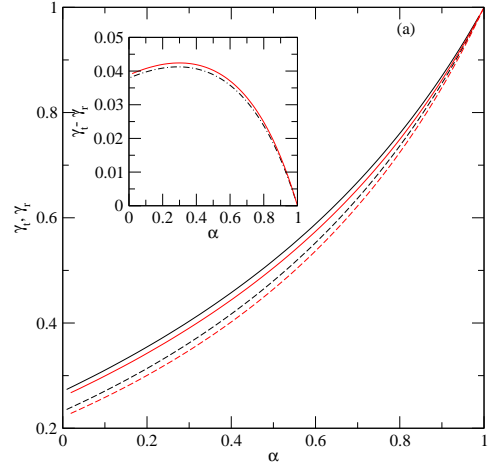}}\\

\resizebox{8.0cm}{!}{\includegraphics{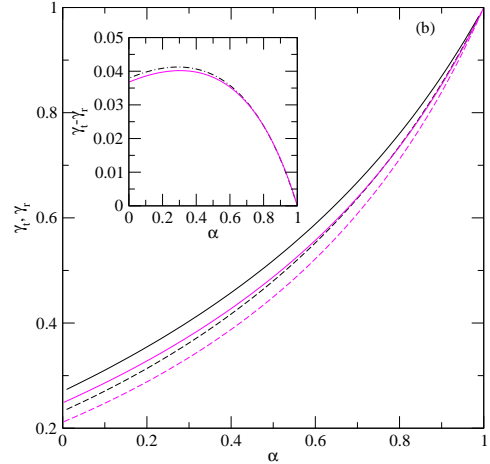}}

\caption{Ratio of the translational (full curve) $\gamma_T$ and
rotational  $\gamma_R$     (dashed  curve)  granular    temperature to  the
temperature of the bath  versus the normal restitution coefficient
$\alpha$ for a system of a needle and a bath of  points  and
a system of a needle and a bath of disks  for $c=2/3$
(a) and for $c=1/10$. The curves corresponding to the needle-point
system are always above those for the needle-disk system (b). The insets display the differences
$\gamma_T-\gamma_R$ versus $\alpha$ for the needle in the bath of disks (full curve)
and the needle in the bath of points (dashed curve).}\label{fig:2}
\end{figure}

After inserting  Eq.~(\ref{eq:15}) in  Eq.~(\ref{eq:14}) it is necessary
to  perform  integrations   over each  variable in  the  corresponding
equation for  the  rotational energy. Details of this  rather  technical  task are
given in the Appendix  B. After some  tedious calculation, one obtains
the following set of equations

\begin{align}\label{eq:18}
b&\left(cI_{1}^{01}(a,k)+(1-c)J_{1}^{01}(a,k)\right)=\nonumber\\
&\frac{1+\alpha}{2}\left(cI_{2}^{03}(a,k)+(1-c)J_{2}^{03}(a,k)\right)\\\label{eq:19}
a&\left(cI_{1}^{11}(a,k)+(1-c)J_{1}^{11}(a,k)\right)=\nonumber\\
&\frac{1+\alpha}{2}\left(cI_{2}^{13}(a,k)+(1-c)J_{2}^{13}(a,k)\right)
\end{align}
where 
\begin{equation}
k=\frac{L^2}{4I\left(\frac{1}{m}+\frac{1}{M}\right)},
\end{equation}
\begin{equation}\label{eq:20}
a=\gamma_R\frac{M+m}{M+m\gamma_T},
\end{equation}
\begin{equation}\label{eq:21}
b=\gamma_T\frac{M+m}{M+m\gamma_T}.
\end{equation}
and 
\begin{equation}\label{eq:22}
c=\frac{L/2}{L/2+(r+R)}
\end{equation}
and $I_{m}^{np}$ and $J_{m}^{np}$ are given by 
\begin{align}
I_{m}^{np}(u,v)=\int_0^1dx\frac{x^{2n}(1+uvx^2)^{p/2}}{(1+vx^2)^m}\\
J_{m}^{np}(u,v)=\int_0^{\frac{\pi}{2}}d\theta 
\sin^{2n} (\theta) \frac{(1+uv\sin^2 \theta)^{p/2}}{(1+v\sin^2 \theta)^m}
\end{align}

Explicit  expressions    for the integrals  appearing  in
Eqs.~(\ref{eq:18})    and    (\ref{eq:19})  are    given   in  Appendix
C. Eq.~(\ref{eq:19}) is an implicit equation for $a$ that, for a given
value of $\alpha$, can  be solved with  standard numerical methods.  $b$ is
then  easily   obtained  by calculating   the   ratio of integrals  of
Eq.~(\ref{eq:18}). Finally,  from the values of $a$  and $b$, $\gamma_T$ and
$\gamma_R$ can be obtained from Eqs.(\ref{eq:20})-(\ref{eq:21}).

A first  check  is  the elastic case, $\alpha=1$, for which one obtains   
$a=b=1$ which  gives $\gamma_T=1$  and,  since $a/b=\gamma_R/\gamma_T$,  $\gamma_R=1$ i.e.,
the temperatures of translational and rotational degrees of
freedom are the same and  correspond to the bath temperature, a
property of an equilibrium system.

In  the  limit $R\to0$  and $r\to0$, for   which $c\to1$  corresponding to a
needle  in  a bath  of point  particles,  one  recovers the results of
Ref\cite{VT04}. More interesting is the limit $R\to0$ with $r$ remaining
finite,  i.e., a  needle  in   a bath   of   {\it  disks}.  By   using
Eq.(\ref{eq:22}) one obtains that  $c=L/(L+2r)$. Since $c<1$, unlike the
simple needle-point system,  there  is a contribution  resulting  from
collisions between bath particles and the  needle's tips. This type of
collision  is  dominant when the  bath  particles are larger  than the
longest size of the anisotropic particle.

 To  illustrate the effect of
this contribution, Figures~\ref{fig:2}a and b  compare the system of a
needle in a  bath of point particles and  the system of  a needle in a
bath of disks,   the radius of  disks $r$  are  equal to  $L/4$  in
Fig~\ref{fig:2}a and $r=9L/2$  in Fig~\ref{fig:2}b.  It is  noticeable
that the translational    and rotational temperatures of  the   latter
system are smaller than  those of the needle  in a bath of  points, an
effect that becomes more  pronounced when  the radius of  the disks  becomes larger
than the length  of the needle.  The difference $\gamma_T-\gamma_R$ are shown in
the insets for  the needle and the bath  of points and for  the needle
and the bath  of disks. Note that  the translational  temperature is
always larger than the rotational temperature  and that the difference
depends  very   weakly  on     the  size  of     the radius  of    the
bath. Additionally,  the  difference  increases when   the restitution
coefficient decreases from $1$ (elastic case), reaches a maximum for a
value of $\alpha\sim    0.3$   and decreases slightly  when   the  restitution
coefficient still decreases.

\begin{figure}[t]
\resizebox{9cm}{!}{\includegraphics{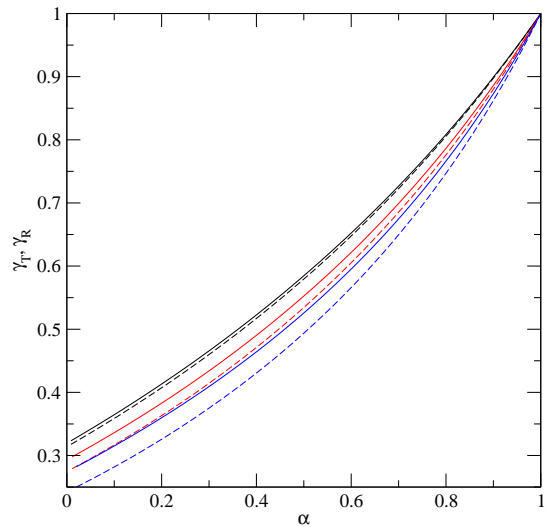}}

\caption{Ratio of the translational (full curve) $\gamma_T$ and
rotational  $\gamma_R$     (dashed  curve)  granular    temperature to  the
temperature of the bath  versus the normal restitution coefficient
$\alpha$ for a discorectangle  and a bath of  disks, for different values
of the anisotropy $r=R$ and $L=R,3R,10R$}\label{fig:3}
\end{figure}

Figure~\ref{fig:3} shows the influence of the anisotropy of the tracer
particle. The rotational temperature is still lower than the translational
temperature what ever the elongation of the particles, but the effect
is greater when the elongation is large. The two upper curves
correspond the translational (full curve) and rotational (dashed
curve) temperature of the
discorectangle when $L=R$, the two intermediate curve to granular
temperatures when $L=3R$, and the two lower curves for a
discorectangle with $L=10R$.

When the anisotropy of the
discorectangle approaches zero, i.e. $L/R\to0$, $c\to0$, one can show by using
Eqs.(\ref{eq:18})-(\ref{eq:19}) that 
\begin{equation}
\gamma_T=\frac{1+\alpha}{2+\frac mM(1-\alpha)}.
\end{equation}
which is the  the result of Martin and 
Piasecki\cite{MP99} for  a spherical tracer particle in a bath of
spherical particles. Moreover, the limit $L/R\to0$
leads equipartition between the rotational and translational granular
temperatures, a result which is different from a model
of pure spherical particles where, in the absence of tangential
friction, the particles cannot exchange rotational
energy. 
In physical systems, the particles are never completely
spherical and our model shows that if an infinitesimal amount of
anisotropy is present, the translational and rotational temperatures
are equal in the steady state. Although our analysis provides no quantitative
information about the relaxation time to reach this NESS, it is
certain to be very long in this limit of small anisotropy.

\subsection{Influence of mass ratio}
\begin{figure} \centering
\resizebox{9cm}{!}{\includegraphics{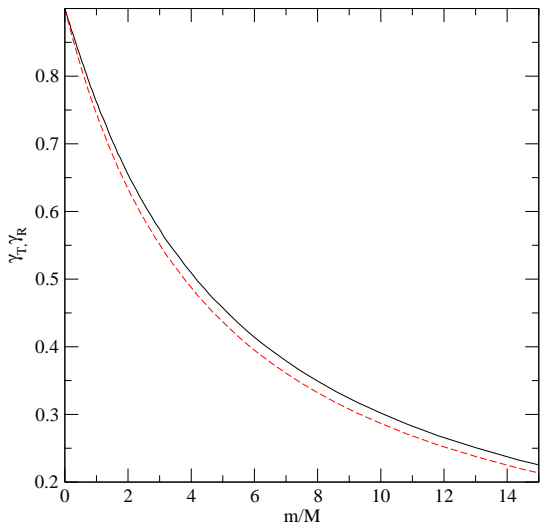}}
\caption{ Ratio of the granular temperatures
 $\gamma_T$ (full curve) 
$\gamma_R$ (dashed curve) to the bath temperature
as a  function of the mass ratio  $m/M$, for a homogeneous discorectangle
with  $c=4/7$.}\label{fig:4}
\end{figure}

We  consider a homogeneous discorectangle of mass $M$ in  a bath of
disks each of mass $m$ for which
$M\neq m$.    When        $m/M\to0$ one obtains,   from
Eqs.(\ref{eq:18})-(\ref{eq:19}), that
\begin{equation}
\gamma_T=\gamma_R=\frac{1+\alpha}{2}
\end{equation}
i.e., equipartition between the degrees of freedom of the tracer
particle, but not between the bath and the tracer particle.  This is,
moreover, the same result for a needle in a bath of point
particles. We conjecture that this result is general in the sense that
we expect equipartition between the different degrees of freedom of
the tracer particle in a bath of light particles, whatever the shape
of the tracer particle and the dimension of the system. The behavior
for finite values of the ratio $m/M$ is shown in Fig.\ref{fig:4}. The
granular temperatures decrease when the ratio $m/M$ increases, and the
translational temperature remains higher than the rotational
temperature for each value of $\alpha$.

\section{Non uniform restitution coefficient}
In practice  it may  be  difficult to construct  a  discorectangle for
which   the  restitution  coefficient   is  constant  over  the entire
perimeter. It is clear, for example, that if the object is composed of
a homogeneous viscoelastic material, collisions with  the ends will be
characterized  by a  smaller  restitution coefficient than  collisions
with  the  linear part.  This effect  becomes  more  pronounced as the
elongation ($L/R$)  increases.  Other    possibilities exist  for    a
non-homogeneous discorectangle composed  of two or more materials. For
example, a hard material may be used to construct the caps. In
addition, the restitution coefficient could depend on the relative
velocity of the point of impact\cite{RPBS99}, an effect that one
neglects here as a first approximation.

\begin{figure}
\centering

\begin{picture}(0,0)%
\includegraphics{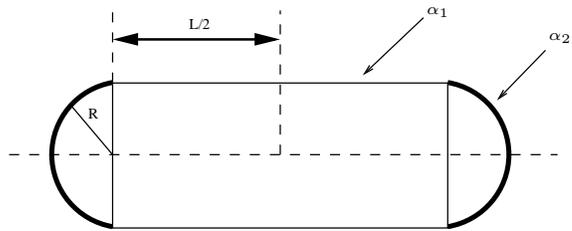}%
\end{picture}%
\setlength{\unitlength}{4144sp}%
\begingroup\makeatletter\ifx\SetFigFont\undefined%
\gdef\SetFigFont#1#2#3#4#5{%
  \reset@font\fontsize{#1}{#2pt}%
  \fontfamily{#3}\fontseries{#4}\fontshape{#5}%
  \selectfont}%
\fi\endgroup%
\begin{picture}(3305,1368)(1159,-1334)
\put(4398,-190){\makebox(0,0)[lb]{\smash{{\SetFigFont{6}{7.2}{\rmdefault}{\mddefault}{\updefault}$\alpha_2$}}}}
\put(3668,-38){\makebox(0,0)[lb]{\smash{{\SetFigFont{6}{7.2}{\rmdefault}{\mddefault}{\updefault}$\alpha_1$}}}}
\end{picture}%

\caption{Sketch of the discorectangle with two restitution
coefficients $\alpha_1$ (rectilinear part) and $\alpha_2$ (circular part).}\label{fig:5}
\end{figure}

As a first approach to describe this possibility, we
consider in this section a discorectangle where the restitution
coefficient is equal to $\alpha_1$ for a collision along the rectilinear
part of the object and equal to $\alpha_2$ for one along the circular part:
see Fig.\ref{fig:5}.
\begin{figure}
\resizebox{9cm}{!}{\includegraphics{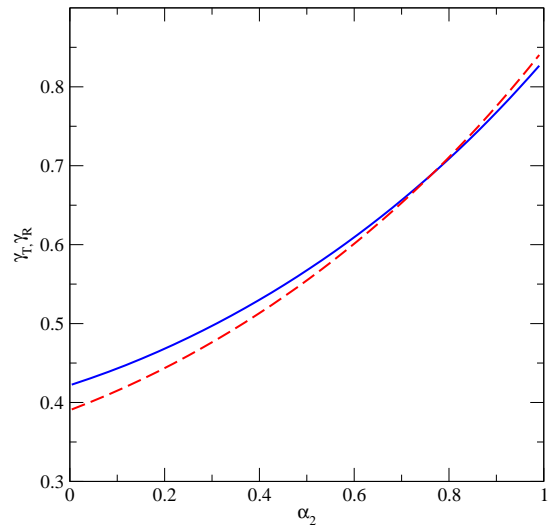}}

\caption{Ratio of the translational (full curve) $\gamma_T$ and
rotational $\gamma_R$    (dashed  curve)   granular  temperature   to   the
temperature of   the bath versus  the  normal  restitution coefficient
$\alpha_2$    with  a normal   restitution  coefficient   $\alpha_1=0.5$  for  a
homogeneous  discorectangle with  $M=m$,  $L=2R$ and $R=r$. Note  that
equipartition    is   recovered   for    a  non-trivial      value  of
$\alpha_2$}\label{fig:6}
\end{figure}
Using the procedure outlined above 
one obtains the following set of closed equations
\begin{align}\label{eq:23}
b&\left((1+\alpha_1)cI_{1}^{01}(a,k)+(1+\alpha_2)(1-c)J_{1}^{01}(a,k)\right)\nonumber\\
&=\frac{(1+\alpha_1)^2}{2}cI_{2}^{03}(a,k)+\frac{(1+\alpha_2)^2}{2}(1-c)J_{2}^{03}(a,k)\\\label{eq:24}
a&\left((1+\alpha_1)cI_{1}^{11}(a,k)+(1+\alpha_2)(1-c)J_{1}^{11}(a,k)\right)\nonumber\\
&=\frac{(1+\alpha_1)^2}{2}cI_{2}^{13}(a,k)+\frac{(1+\alpha_2)^2}{2}(1-c)J_{2}^{13}(a,k)
\end{align}
Figure~\ref{fig:6} shows  the  ratio of  translational  and rotational
temperature of  a    discorectangle with  $R=r$   and  $L=2R$  to  the
temperature  of the bath as   a  function of  the normal   restitution
coefficient  $\alpha_2$  with a    fixed  $\alpha_1=0.5$.  One notes   that  the
translational    temperature   becomes smaller    than  the rotational
temperature for $\alpha_2>0.765$.  When the two curves cross, equipartition
is  recovered, but unlike   the limiting cases   discussed above  of a
discorectangle  with  a uniform   restitution coefficient (light  bath
particles, infinitely  small anisotropy),  for a non-trivial  value of
restitution coefficient $\alpha_2$  and,   moreover, for larger  values  of
$\alpha_2$ the ratio of temperatures is inverted.

One can determine in general 
when equipartition is recovered for a discorectangle with two
restitution coefficients. Using the relation $a=b$ (assumption of equipartition)
and Eqs.(\ref{eq:23})-~(\ref{eq:24}), one obtains two  implicit equations
with the three parameters $\alpha_1$, $\alpha_2$ and $a$. A simple numerical
procedure allows us to obtain the $\alpha_2$ as a function of $\alpha_1$.

\begin{figure}
\resizebox{9cm}{!}{\includegraphics{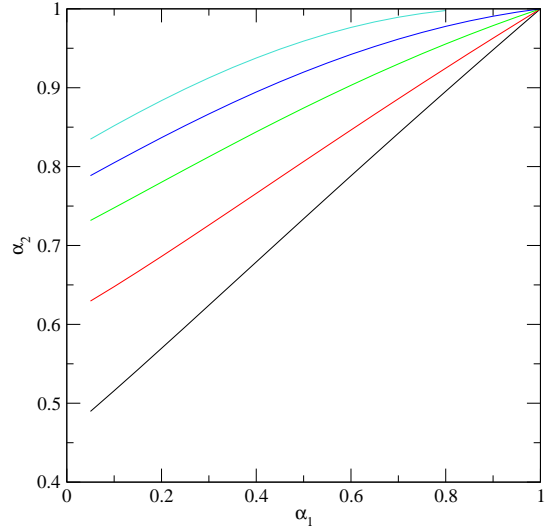}}

\caption{$\alpha_1$ versus $\alpha_2$ where equipartition is obtained
for $r=R$ and different values of $L=R,2R,\ldots,5R$ from
bottom to top}
\label{fig:7}
\end{figure}

Figure~\ref{fig:7} shows the equipartition lines in the $(\alpha_1,\alpha_2)$
space. Above each line, corresponding to a given elongation of the
rectangle, $T_R>T_T$ while the reverse inequality applies to the
region below the line.  
It is noticeable that as the elongation increases,  the region of the $(\alpha_1,\alpha_2)$
space where $T_R>T_T$ decreases.

\begin{figure}
\resizebox{9cm}{!}{\includegraphics{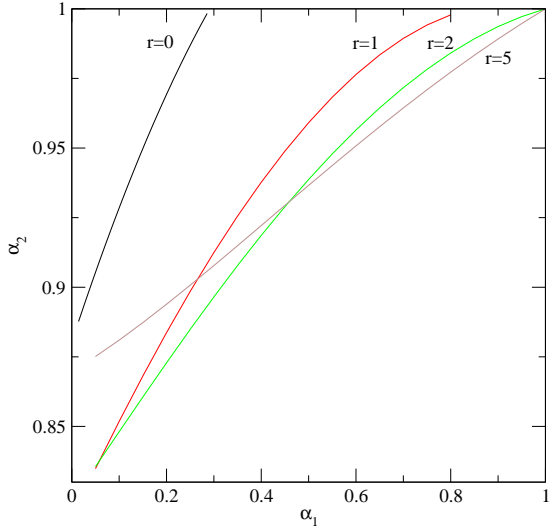}}

\caption{$\alpha_1$ versus $\alpha_2$ where equipartition is obtained for $L=8R$ 
and  for different  values  of  disk radius  $r/R=0,  1,2,3,5$ (number
labelling each curve)}
\label{fig:8}
\end{figure}

Figure~\ref{fig:8} shows the role of the size of the bath particles on
the existence of equipartition  of a  discorectangle  of length
$L=8R$. For small bath particles, only a small range of $\alpha_1$, (
between $0$ and $\sim 0.3$) with a smaller range of $\alpha_2$ (
between $\sim0.89$ and $1$) allows equipartition for the tracer
particle. For larger bath disks, all values
of $\alpha_1$ (between $0$ and $1$) are available with a smaller
corresponding range of $\alpha_2$.

\section{Conclusion}
We  have investigated  the influence of   the anisotropy  of a  tracer
particle in a bath of thermalized disks  in two dimensions. By using a
mean-field  approach,  we have   obtained analytical  results  for the
rotational  and   translational    temperatures. For   a   homogeneous
discorectangle with  a  uniform  normal restitution  coefficient,  the
translational temperature  is   always  higher   than  the  rotational
temperature, with the  difference  depending on the elongation  of the
tracer particle, the size of the bath particles and the mass ratio.

At present, the exact dependence of the restitution coefficient on the
position of the point of impact  is unknown. It seems likely, however,
that it is not constant along the perimeter. As a first approximation,
we considered an extension of the model where it  takes on two values,
one for collisions with  the  linear part  and another for  collisions
with the  caps. In this case the  difference between the translational
and  rotational granular  temperatures   can  be either  positive   or
negative  depending on the values of  the restitution coefficients and
the   other parameters. It is   clearly   possible to generalize   the
calculation   to allow for    a   continuously  varying    restitution
coefficient.

It would also
be interesting to investigate a system of a tracer
particle with a small anisotropy where the tangential restitution
coefficient 
has a non-trivial value ($-1<\alpha_T<1$) where the limits correspond
to a perfectly rough and a perfectly smooth surface, respectively. The
intermediate situation corresponds to some friction which is important
in  some granular systems. 

\appendix 
\section{Moment of Inertia }
For a homogeneous discorectangle, the moment of inertia is given by
\begin{equation}
I_{Oz}=\int\int_S\rho (x^2+y^2)dxdy
\end{equation}
and the mass of the system is
\begin{equation}
M=\int\int_S \rho dxdy=\rho(\pi R^2+2LR).
\end{equation}
which gives 
\begin{equation}
I_{Oz}=\rho R\Big[\pi R\Big(\frac {R^2}2+\Big(\frac
L2\Big)^2\Big)+\frac23L\Big(3R^2+\Big(\frac L2\Big)^2\Big)\Big].
\end{equation}
By substituting the density as a function of the total mass of the discorectangle.
\begin{equation}
I=M\left[\frac{\frac{2L\left(3R^2+\left(\frac{L}{2}\right)^2\right)}{3}+\pi
R \left(\frac{R^2}{2}+\left(\frac{L}{2}\right)^2\right)}{\pi R +2L}\right]
\end{equation}

Two well-known limits are recovered
\begin{align}
\lim_{R\to0}I_{Oz} &=\frac{ML^2}{12} \\
\lim_{L\to0}I_{Oz} &=\frac{MR^2}{2}
\end{align}
For an inhomogeneous discorectangle, the moment of inertia depends on
the mass distribution. However, it is possible to determine the lower
and upper bounds for allowable values. A trivial lower bound of zero
is obtained when the mass is concentrated at the center. Conversely,
when the mass is distributed equally at the two extremities of the
object (point masses of $M/2$ at a distance of $L+R$ from the center
on each side), one obtains
\begin{equation}
I=M(L/2+R)^2
\end{equation}
which gives the upper bound for the moment of inertia.
 
\section{Needle average energy loss} As for binary mixtures
of spheres\cite{BT02}, we use  a Gaussian ansatz for  the distribution
functions and  introduce two  different temperatures corresponding  to
the  translational and  rotational degrees  of  freedom of the needle.
The homogeneous distribution functions of the needle and of the points
are then  given respectively by 
\begin{equation} f({\bf v}_1,\omega_1)\sim
\exp\left(-\frac{M{\bf v}_1^2\gamma_T^{-1}}{2T}-\frac{I\omega_1^2\gamma_R^{-1}}{2T}\right),
\end{equation}

\begin{equation} \Phi({\bf v}_2)\sim
\exp\left(-\frac{m{\bf v}_2^2}{2T}\right),
 \end{equation}
where $T$ is the temperature of the bath, $\gamma_T$ and $\gamma_R$
the ratio of the translational (and rotational) temperature of the
needle to the bath temperature.

We introduce the vectors ${\boldsymbol \chi}$ and ${\boldsymbol \nu}$ such that
\begin{align}\label{eq:25}
{\boldsymbol \chi}&=\frac{1}{\sqrt{2T(M\gamma_T+m)}}\left(M{\bf v}_1+m{\bf v}_2\right)\\
{\boldsymbol \nu}&=\sqrt{\frac{mM}{2T(M\gamma_T+m)\gamma_T}}\left({\bf v}_1-\gamma_T{\bf v}_2\right)
\end{align}
The scalar products ${\bf V}.{\bf u}_1^\perp $ and ${\bf V}.{\bf u}_r$ can be expressed as
\begin{align}\label{eq:26}
{\bf V}.{\bf u}_1^\perp &=h\Big[(\gamma_T-1){\bf \chi}.{\bf
u}_1^\perp+\sqrt{\gamma_T}\Big(\sqrt{\frac{m}{M}}+\sqrt{\frac{M}{m}}\Big){\boldsymbol
\nu}.{\bf u}_1^\perp\Big]\nonumber\\
&+\omega_1\lambda \\\label{eq:27}
{\bf V}.{\bf u}_r &=h\Big[(\gamma_T-1){\bf \chi}.{\bf
u}_r+\sqrt{\gamma_T}\Big(\sqrt{\frac{m}{M}}+\sqrt{\frac{M}{m}}\Big){\boldsymbol
\nu}.{\bf u}_r\Big]\nonumber\\
&+\omega_1\frac{L}{2}\sin\theta
\end{align}
where $h=\sqrt{\frac{2T}{M\gamma_T+m}}$.

Let us introduce $\xi=\omega_1\sqrt{\frac{I}{2T\gamma_R}}$.
The translational energy loss is given by the formula
\begin{align}\label{eq:28}
&\sum_{p=\pm1}\bigg[\int d\lambda\int \frac{d\theta_1}{2\pi}\int d{\bf \chi}\int d{\bf \nu}\int d\xi\exp(-{\bf
\chi}^2-{\bf \nu}^2-\xi^2)
\nonumber\\&|{\bf V}.{\bf u}_1^\perp|\Theta(p{\bf V}.{\bf u}_1^\perp)\Theta(\frac{L}{2}-|\lambda|)\Delta E_1^T + \nonumber \\
&(R+r)\int d\theta\int \frac{d\theta_1}{2\pi}\int d{\bf \chi}\int d{\bf \nu}\int d\xi\exp(-{\bf
\chi}^2-{\bf \nu}^2-\xi^2)|\nonumber \\&
{\bf V}.{\bf u}_r|\Theta(p{\bf V}.{\bf u}_r)\Delta E_1^T\bigg] =0
\end{align}

Since Eq.~(\ref{eq:26})  depends only on ${\boldsymbol \chi}.{\bf u}_1^\perp$
and
${\boldsymbol \nu}.{\bf u}_1^\perp$ , one can freely integrate over the
direction of ${\bf u}_1$ for the vectors ${\boldsymbol \chi}$ and
${\boldsymbol \nu}$, and similarly for Eq.(\ref{eq:27}). The
integration over $\theta_1$ can be easily performed. If we introduce the
three dimensional vectors ${\bf G}_{{\bf u}_1^\perp}$ and ${\bf s}_{{\bf
u}_1^\perp}$ with components:
\begin{align}
{\bf G}_{{\bf u}_1^\perp} &=(G_1,G_2,G_3) \\   &=\Big(\sqrt{\frac{2T}{M\gamma_T+m}}(\gamma_T-1),\sqrt{\frac{2T\gamma_T}{M\gamma_T+m}}\frac{m+M}{\sqrt{mM}},\lambda\sqrt{\frac{2T\gamma_R}{I}}\Big)
\end{align}
and 
\begin{align}
{\bf s}_{{\bf u}_1^\perp} &=(s_1,s_2,s_3) \\
        &=({\bf \chi}.{\bf u}_1^\perp,{\bf \nu}.{\bf u}_1^\perp,{\xi})
\end{align}
Respectively, one has introduces  ${\bf
G}_{{\bf u}_r}$ and ${\bf s}_{{\bf u}_r}$ vectors associated with
collisions on the circular parts of the discorectangle
by changing  ${\bf u}_1^\perp$ in ${\bf
u}_r$ and $\lambda$ by  $\frac L2 \sin\theta$.
By inserting Eq.~(\ref{eq:16}) in Eq.~(\ref{eq:28}), the  average energy
loss  can be rewritten as

\begin{align}\label{eq:29}
&\sum_{p=\pm1}\int_{-L/2}^{L/2}     d\lambda\int        d{\bf      s}_{u_1^\perp}\exp(-{\bf
s}_{u_1^\perp}^2)|{\bf G}_{u_1^\perp}.{\bf s}_{u_1^\perp}|\Theta(p{\bf  G}_{u_1^\perp}.{\bf
s}_{u_1^\perp})\nonumber\\           &\bigg[\frac{1}{2M}\frac{(1+\alpha)^2({\bf
G}_{u_1^\perp}.{\bf
s}_{u_1^\perp})^2}{(\frac1m+\frac1M+\frac{\lambda^2}{I})^2}-\nonumber         \\
&\frac{(1+\alpha){\bf                                       G}_{u_1^\perp}.{\bf
s}_{u_1^\perp}}{\frac1m+\frac1M+\frac{\lambda^2}{I}}\sqrt{\frac{2T}{M\gamma_T+m}}\Big(\gamma_Ts_1+\sqrt{\frac{m\gamma_T}{M}}s_2\Big)\bigg]\nonumber
\\  &+   (R+r)\int_0^{2\pi} d\theta\int  d{\bf s}_{u_r}\exp(-{\bf  s}_{u_r}^2)|{\bf
G}_{u_r}.{\bf s}_{u_r}|\Theta(p{\bf G}_{u_r}.{\bf s}_{u_r})\nonumber\\&
\bigg[\frac{1}{2M}\frac{(1+\alpha)^2({\bf G}_{u_r}.{\bf s}_{u_r})^2}{(\frac1m+\frac1M+\frac{L^2\sin^2\theta}{4I})^2}\nonumber \\
&-\frac{(1+\alpha){\bf                                        G}_{u_r}.{\bf
s}_{u_r}}{\frac1m+\frac1M+\frac{L^2\sin^2\theta}{4I}}\sqrt{\frac{2T}{M\gamma_T+m}}\Big(\gamma_Ts_1+\sqrt{\frac{m\gamma_T}{M}}s_2\Big)\bigg]=0.
\end{align}

By defining a new coordinate system in which  the $z$-axis is parallel
to ${\bf G}$, one find that the 
integrals of Eq.~(\ref{eq:29}) involve  gaussian integrals of the form
\begin{align}
\int d{\bf s}\exp\left(-{\bf s}^2\right)
(|{\bf G}|s_z)^2 \Theta (\pm s_z)G_is_z=\frac{\pi}{2}|{\bf G}|^2G_i
\end{align}
and 
\begin{align}
\int d{\bf s}\exp\left(-{\bf s}^2\right)(|{\bf G}|s_z)^3
 \Theta (\pm s_z)=\frac{\pi}{2}|{\bf G^3}|
\end{align}
which finally leads to  Eq.~(\ref{eq:18}). The equation for  rotational
energy is derived following exactly the same procedure.
\section{Integrals}
The coupled equations (\ref{eq:18})-~(\ref{eq:19}) depend on the eight
integrals $I_{1}^{01}(a,k)$, $I_{2}^{03}(a,k)$, $I_{1}^{11}(a,k)$, $I_{2}^{13}(a,k)$\cite{Note1},
$J_{1}^{01}(a,k)$, $J_{2}^{03}(a,k)$,  $J_{1}^{11}(a,k)$, $J_{2}^{13}(a,k)$ which can be
expressed in terms of transcendental and special functions.
For completeness, we give below their expressions.
\begin{align}
I_{1}^{01}(a,k)=&\sqrt{\frac{a}{k}}\ln(\sqrt{ak}+\sqrt{1+ak})\nonumber\\&+\sqrt{\frac{1-a}{k}}\arctan\left(\sqrt{\frac{(1-a)k}{1+ak}}\right)
\end{align}

\begin{align}
I_{2}^{03}(a,k)&=\frac{a^{3/2}}{\sqrt{k}}\ln(\sqrt{ak}+\sqrt{1+ak})+\frac{(1-a)\sqrt{1+ak}}{2(1+k)}\nonumber\\
&+\frac{1+a-2a^2}{2\sqrt{k(1-a)}}\arctan\left(\sqrt{\frac{(1-a)k}{1+ak}}\right)
\end{align}
\begin{align}
I_{1}^{11}(a,k)
&=\frac{\sqrt{ak+1}}{2k}+\frac{1-2a}{2\sqrt{a}k^{3/2}}+\ln(\sqrt{ak}+\sqrt{1+ak})\nonumber\\&-\frac{\sqrt{1-a}}{k^{3/2}}\arctan\left(\sqrt{\frac{(1-a)k}{1+ak}}\right)
\end{align}
\begin{align}
I_{2}^{13}(a,k)&=
\frac{\sqrt{1+ak}(ak-1+2a)}{2k(k+1)}\nonumber\\
&+\frac{\sqrt{a}(3-4a)}{2k^{3/2}}
\ln(\sqrt{ak}+\sqrt{1+ak})\nonumber\\
&+\frac{\sqrt{1-a}(1-4a)}{2k^{3/2}}\arctan\left(\sqrt{\frac{(1-a)k}{1+ak}}\right)
\end{align}

\begin{align}
J_{1}^{01}(a,k)=&a K(\sqrt{-ak})+(1-a)\Pi(-k,\sqrt{-ak})
\end{align}
where $K(x)$ and $\Pi(x,y)$ denote the complete elliptic integral of
the first kind and the incomplete elliptic integral of the third kind
respectively.
\begin{align}
J_{2}^{03}(a,k)=&\left(a^2-\frac{(1-a)^2}{2(k+1)}\right)\frac{K(\sqrt{\frac{ak}{(1+ak)}})}{\sqrt{(1+ak)}}
\nonumber\\
&+\frac{(1-a)\sqrt{(1+ak)}}{2(k+1)}E(\sqrt{\frac{ak}{1+ak}})
\nonumber\\
&+\frac{(1-a)(2ak+a+k+2)}{2(k+1)^2\sqrt{(1+ak)}}\Pi(\frac{k}{k+1},\sqrt{\frac{ak}{1+ak}})
\end{align}
where $E(x)$ denotes the complete elliptic integral of the
second kind.
\begin{align}
J_{1}^{11}(a,k)=&-\frac{a}{k\sqrt{(1+ak)}}K(\sqrt{\frac{ak}{1+ak}})\nonumber\\
&+\frac{\sqrt{1+ak}}{k}
E(\sqrt{\frac{ak}{1+ak}})\nonumber\\
&-\frac{1-a}{k(k+1)\sqrt{(1+ak)}}\Pi\left(\frac{k}{k+1},\sqrt{\frac{ak}{1+ak}}\right)
\end{align}
and finally
\begin{align}
J_{2}^{13}(a,k)&=
\frac{1+2ak-3a^2-4a^2k}{2k(k+1)\sqrt{1+ak}}K(\sqrt{\frac{ak}{1+ak}})\nonumber\\
&+\frac{\sqrt{1+ak}(2ak+3a-1)}{2k(k+1)}E(\sqrt{\frac{ak}{1+ak}})\nonumber\\
&+\frac{k+4a^2k-3a+3a^2-5ak}{2k(k+1)^2\sqrt{(1+ak)}}\Pi\left(\frac{k}{k+1},\sqrt{\frac{ak}{1+ak}}\right)
\end{align}
%\bibliography{cooling}

\end{document}